\documentclass[a4paper,fleqn,usenatbib]{mnras}

\usepackage[T1]{fontenc}
\usepackage{ae,aecompl}

\usepackage{graphicx}	
\usepackage{amsmath}	
\usepackage{amssymb}	
\usepackage[off]{pstricks}
\usepackage{lineno}
\usepackage{epsfig}   
\usepackage{natbib}  
\usepackage{endnotes}
\usepackage{color}
\usepackage{fixfoot}

\newcommand{\asat}{{\it AstroSat}}

\title[{Time-resolved spectroscopy on the heartbeat state}]{Time-resolved spectroscopy on the heartbeat state of GRS 1915+105 using \asat }

\author[Rawat et. al.]{
Divya Rawat,$^{1,2}$\thanks{E-mail: rawatdivya838@gmail.com (DR)}
Ranjeev Misra,$^{2}$
Pankaj Jain$^{1}$
and J. S. Yadav$^{1}$
\\
$^{1}$Department of physics, IIT Kanpur, Kanpur Nagar, Uttar Pradesh 208016, India\\
$^{2}$Inter-University Center for Astronomy and Astrophysics, Ganeshkhind, Pune 411007, India\\
}

\date{Accepted XXX. Received YYY; in original form ZZZ}

\begin{document}
\label{firstpage}
\pagerange{\pageref{firstpage}--\pageref{lastpage}}
\maketitle
\begin{abstract}
AstroSat spectra of the black hole system GRS 1915+105 during the heartbeat state (with a varying oscillation period from 150 to 100 secs) were analysed using a truncated relativistic disc model along with a Comptonization component. Spectra were fitted for segments of length $\sim 24$ secs. The oscillation can be described as coordinated variations of the accretion rate, Comptonised flux, and the inner disc radius, with the latter ranging from $1.235$-$5$ gravitational radii. 
Comparison with results from the $\chi$ and Intermediate states shows that while the accretion rate and the high energy photon index  were similar, the inner disc radius and the fraction of Comptonised photons were larger for these states than for the heartbeat one.
The coronal efficiency $\eta \equiv L_{ac}/\dot M c^2$, where $L_{ac}$ is the radiative luminosity generated in the corona is found to be approximately $\propto \dot M^{-2/3}$ for all the observations. The efficiency decreases with inner radii for the heartbeat state but has similar values for the $\chi$ and Intermediate states where the inner radii is larger. The implications of these results are discussed.
\end{abstract}

\section{Introduction}
GRS 1915+105 is a galactic black hole binary system with luminosity close to the Eddington limit. A qualitative  behaviour of the transient system in the Hardness Intensity Diagram (HID) \citep{da14} shows that this source executes transitions between hard and soft state via an intermediate state. This source was in outburst since it was discovered in 1992 and went to the obscured state in 2019 with interrupted X-ray flares \citep{ne20}. The binary system contains a black hole of 12.4 solar mass \citep{re14}, placed at a distance of 8.6 kpc \citep{re14} in the Sagittarius arm of the Milky Way galaxy. The companion is a regular K-type star \citep{Vi02} with a binary orbital period of 34 days \citep{gr01}. In the binary system, the compact object is a rapidly rotating Kerr black hole with a dimensionless spin parameter $\geq$0.98 \citep{mc06,bl09,mi20,sr20,ho20}. 

\citet{gr96} first reported a drastic change in X-ray variability on time scale of secs to days. This source shows complex, but nearly periodic and repetitive X-ray variability. The source is divided into 14 different X-ray classes on the basis of its X-ray flux and Color-Color Diagram (CCD) \citep{be00,ke02,ha03}. The long RXTE observations reveal that the source spends most of the time in the hard state, which belongs to $\chi$ class \citep{be00}. In this state, the source does not show much variability. It also spends a significant amount of time in a highly variable $\rho$ class,  also known as the heartbeat class (heartbeat class hereafter), which has been widely studied in the literature \citep{ta97,pa98,ya99,ni11,ne12,pa14,sh18}.  The phase-resolved spectroscopy  for the heartbeat class was done by \citet{ne12} using RXTE/PCA data in 3.3-45.0 keV band. They have explained the heartbeat oscillation as a combination of a thermal–viscous radiation pressure instability and Radiation pressure-driven evaporation or ejection event in the inner accretion disc.

\citet{mi03} performed time-resolved spectroscopy (with each segment of length $\sim$16 sec) on the $\beta$ class using RXTE data in 3.0-25.0 keV band. They studied the variation of $\Gamma$ and $R_{in}$ with time and reported variations of local accretion rate and mass loss.
With AstroSat data, \citet{ra19} have studied the temporal properties of GRS 1915+105 and observed a state transition from $\chi$ class to Heartbeat state via intermediate class (IMS onwards) for 28$^{th}$ March 2017 observation. These transitions occur at unabsorbed 0.1-80.0 keV luminosity $\sim 0.1 L_{edd}$. Next, \citet{mi20} studied the spectro-temporal properties for both 28$^{th}$ March and 1$^{st}$ April 2017 observations and identified the origin of C-type QPO frequency as dynamical frequency. Using the same set of observations as analysed previously by \citet{ra19,mi20}, in this work, we have undertaken time-resolved spectroscopy in the 1.0-50.0 keV band using AstroSat observation for a heartbeat oscillation of GRS 1915+105, to quantify the evolution of the spectral properties during the oscillation. This is in contrast to earlier analysis of RXTE data where a phase resolved spectroscopy method was used \citep{ni11,ne12}. The advantage of  time-resolved spectroscopy is that the analysis is not affected by the selection of the peak of an individual cycle, averaging over cycles with different time-periods and  variation of the profile for different cycles. Thus, the technique provides a more direct approach to understanding spectral evolution. However, the disadvantage of time-resolved spectroscopy is that the length of the time segment used for spectral fitting  (in this case $\sim 24$ seconds) has to be large enough to obtain statistically significant data. Hence, faster spectral variations are not captured. \\
We aim to study the spectral transition of the source and correlation between spectral parameters with simultaneous Soft X-ray Telescope (SXT) and Large Area X-ray Proportional Counters (LAXPCs) data.  Because of the low energy coverage of SXT (0.3-8.0 KeV) in which the disc emission dominates, it enables us to get a reasonable estimate of inner disc properties (like inner radii) which was earlier not possible with RXTE. We have discussed the data analysis techniques in Section 2 followed  by results  in Section 3. The discussion and summary of interpretations are given in Section 4 and 5.
\section{Observation and  Data Reduction}
The galactic micro-quasar source GRS 1915+105 was observed from 28$^{th}$ March 18:03:19 till 29$^{th}$ March 2017 19:54:07 (hereafter epoch 1) and 1$^{st}$ April 00:03:21 till 1$^{st}$ April 14:38:01 2017 (hereafter epoch 2) with SXT \citep{si16, si17} $\&$ LAXPC \citep{ya16a} onboard AstroSat \citep{ag06,si14} satellite. SXT is an imaging instrument that works in 0.3-8.0 keV with a spectral resolution of 90 eV at 1.5 keV. LAXPC is a photon-counting array  and has an effective area of about 6100 cm$^{2}$ (combined 3 units LX10, LX20 and LX30) at 10.0 keV \citep{ya16b}. We used LAXPC and SXT instruments for spectral analysis to cover the broad energy range of 1.0-50.0 keV. The lower energy range $<$1.0 keV is ignored because of uncertainty in effective area and response in the SXT instrument. The high energy range i.e $>$50.0 keV is ignored because of low signal-to-noise ratio. LaxpcSoft \footnote{\label{note1}\url{http://astrosat-ssc.iucaa.in/?q=laxpcData}} software is used to extract the LAXPC spectra in 4.0-50.0 keV band. We have used recent response and background files provided by LAXPC POC team\footnote{\label{note2}\url{https://www.tifr.res.in/~astrosat_laxpc/LaxpcSoft.html}}. We have used only LX10 $\&$ LX20 for the analysis as there was a gas leakage and consequently instability in the gain of LX30. The SXT spectra in 1.0-5.0 keV energy band were generated in XSELECT V2.4j using a merged event file extracted using SXT Event Merger Tool\footnote{\label{note3}\url{https://www.tifr.res.in/~astrosat_sxt/dataanalysis.html}} taking a source region of 12'. The corresponding auxiliary response file $\&$ response matrix file were generated using ARF generation tool\textsuperscript{\ref{note3}} for each epoch. The simultaneous SXT and LAXPC spectra were fitted in {\it{XSPEC version 12.11.0}}.\\
\begin{figure*}
\centering\includegraphics[scale=0.31]{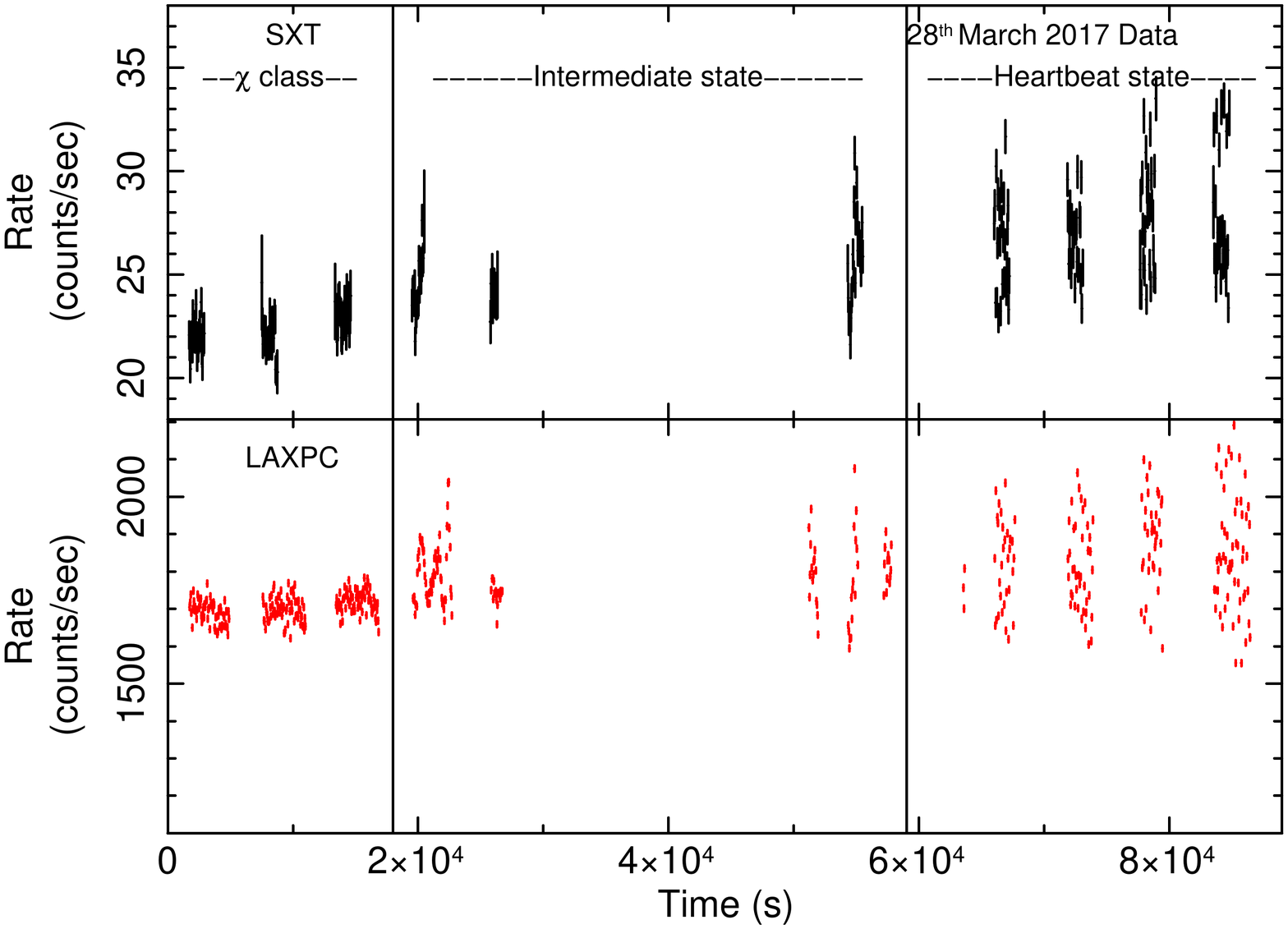}
\centering\includegraphics[scale=0.31]{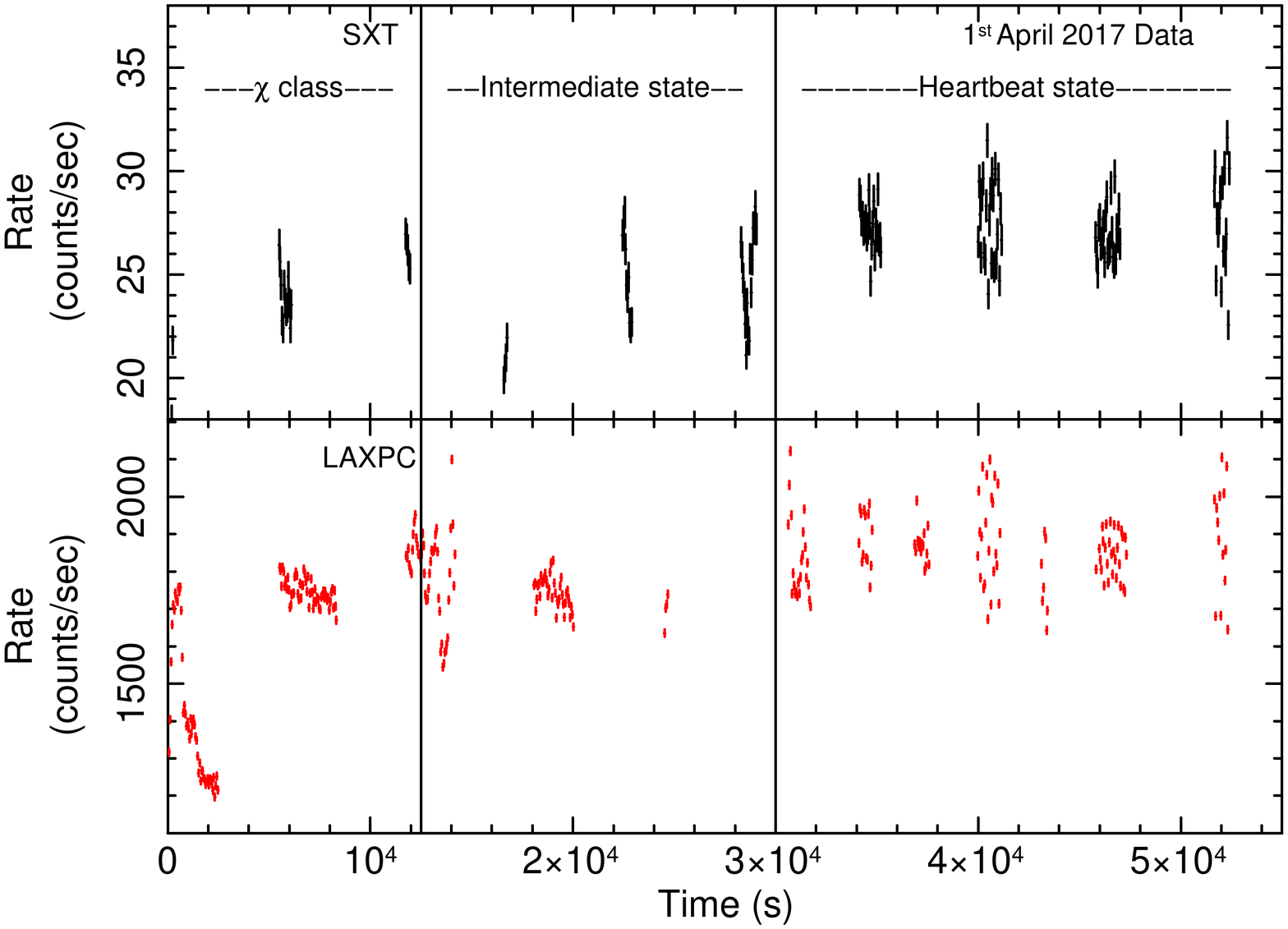}
\caption{The left and right panel shows a full 50 sec binned simultaneous SXT and LAXPC light curve of GRS 1915+105 for epoch 1 and epoch 2, respectively. 
Light curves are plotted in 0.3-8.0 keV (using SXT) and 3.0-80.0 keV (combining {\tt LXP10}, {\tt LXP20}) energy range.}
\label{lightcurve1}
\end{figure*}
\subsection{Spectral Analysis}
\citet{mi20} divided epoch 1 Data into 10 segments and epoch 2 Data into 6 segments (spectral details of each segment is given in Table 1 of \citet{mi20}). Details of the response and background spectra generation for LAXPC can be found in \citet{An17}. To show the extent of the overlap of SXT and LAXPC data for epoch 1 and epoch 2 we have plotted the full SXT and LAXPC light curve in figure \ref{lightcurve1} while \citet{ra19} has shown LAXPC light curve for epoch 1 only. The zoomed view of epoch 1 $\&$ 2 light curves are shown in figure 3, 5 and 7 of \citet{ra19} and figure 1 of \citet{mi20}. The soft count rate in both LAXPC and SXT increases with time as the source transits from $\chi$ class to Heartbeat state. The time-averaged fitted spectra for $\chi$ class, IMS, and Heartbeat state are shown in figure 3 of \citet{mi20}. In the heartbeat state, the X-ray flux shows periodic variation. To study the variation of the spectral parameters over the oscillatory period, we have performed time-resolved spectroscopy on this state.
\begin{figure*}
\includegraphics[scale=0.42,angle=0]{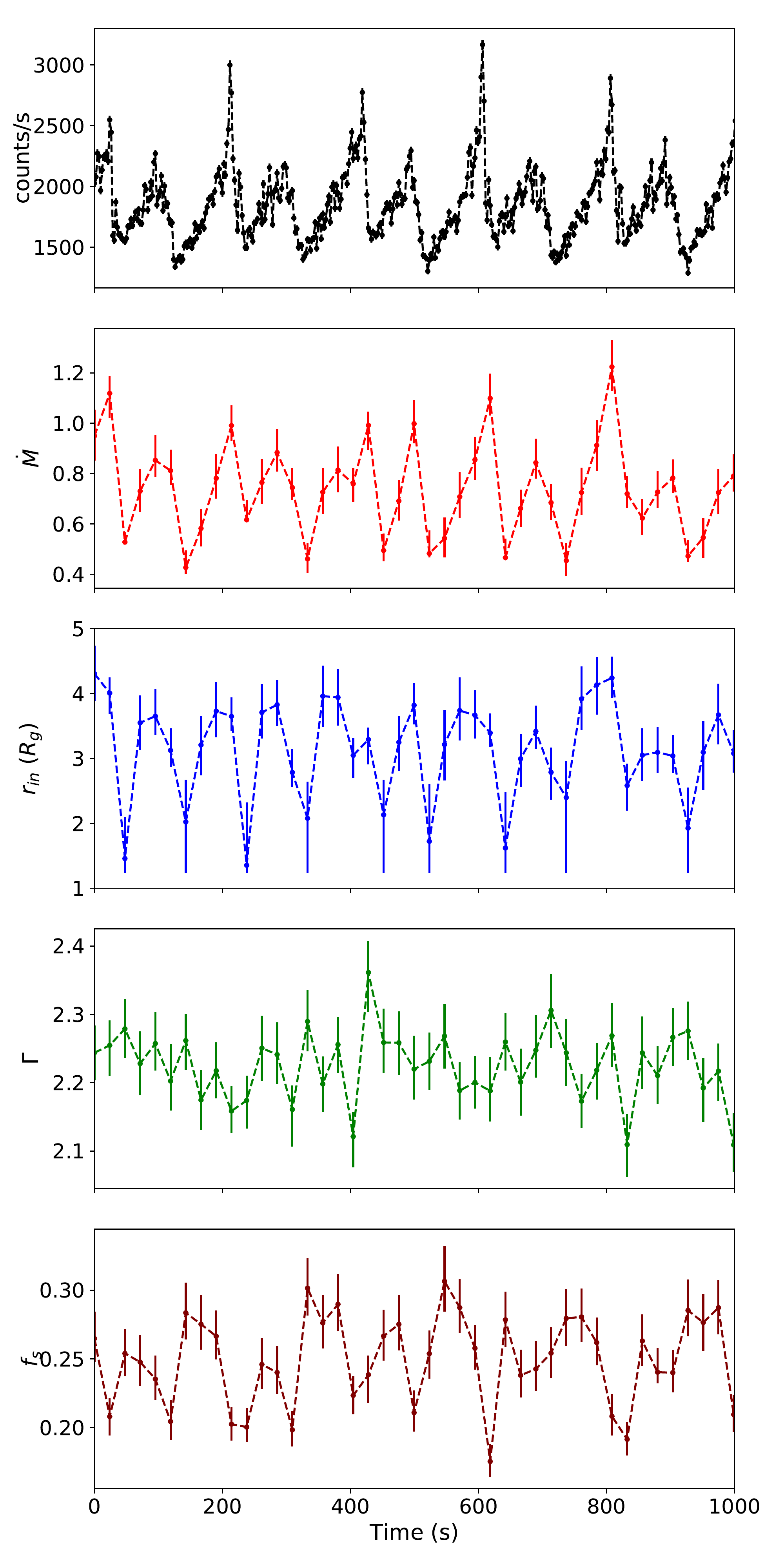}
\includegraphics[scale=0.42,angle=0]{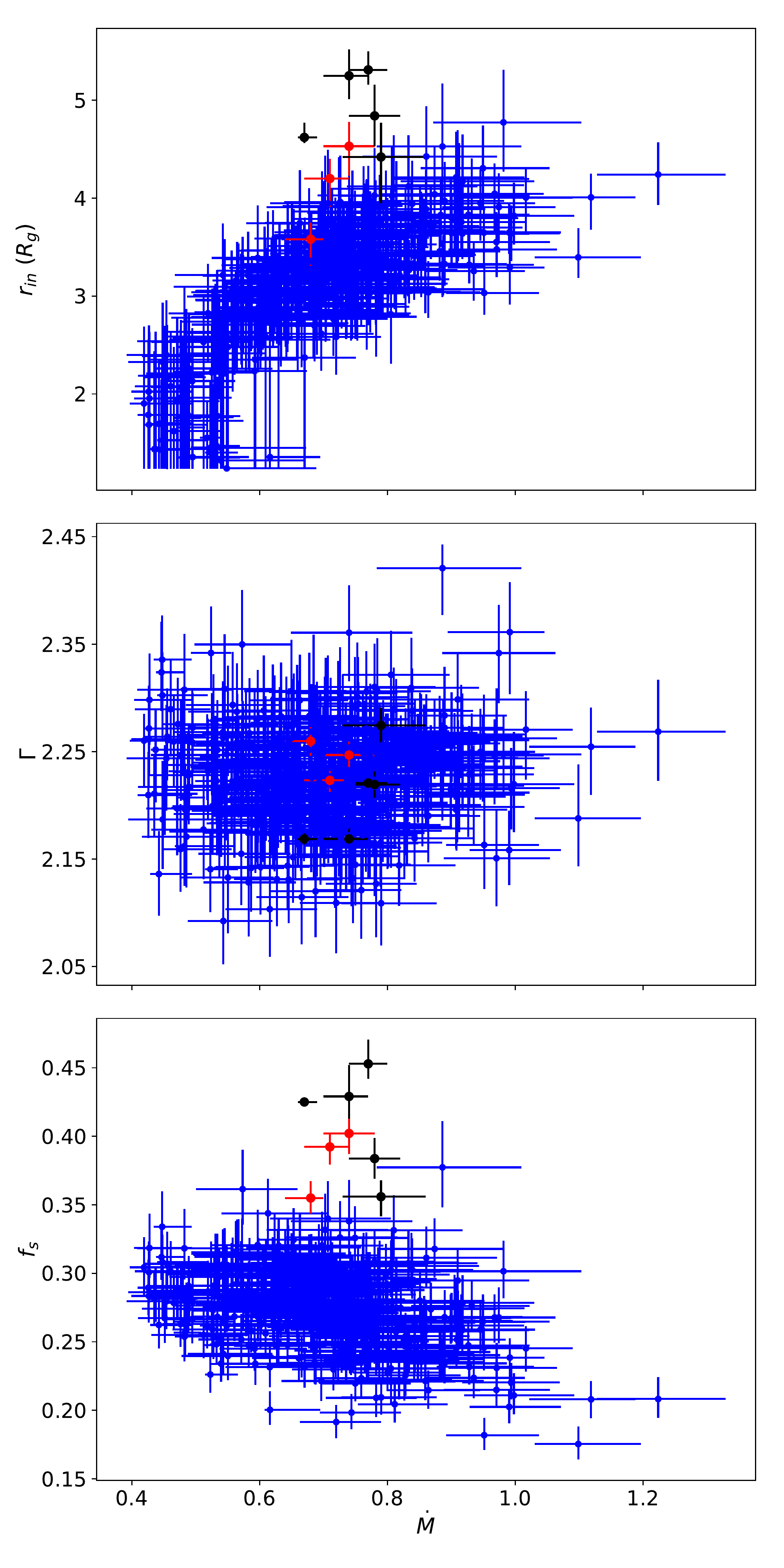}
\caption{The top left panel shows a $\sim$1000 sec light curve, variation of $\dot M$ ($10^{18}gm$ $s^{-1}$), $r_{in}$, $\Gamma$, and $f_{s}$ with time for epoch 1. Here light curve is plotted in 3.0-50.0 keV energy band. The right panel shows the variation of $r_{in}$, $\Gamma$, and $f_{s}$ with $\dot M$ for $\chi$ (black coloured data points), IMS (red coloured data points) and heartbeat state (blue coloured data points) of epoch 1. 
Here, IMS and HS stands for Intermediate state and Heartbeat state respectively}
\label{parm}
\end{figure*}
\subsection{Time-resolved Spectroscopy for Heartbeat state}
We have divided the heartbeat state data into different segments, with each segment equal to 23.775 secs, which is ten times the time resolution of the SXT instrument. The LAXPC and SXT spectra for each segment was extracted. The same model that was used by \citet{mi20} was used to fit the spectra, which is an absorbed thermal comptonization of a relativistic disc represented by the XSPEC model {\it tbabs*(simpl*kerrd)}. Following \citet{mi20}, the absorption column density, the relative constant between the two instruments, and the gain offset for the SXT spectrum were kept fixed to the values obtained from the time-averaged spectra. For all the segments, the reduced $\chi^{2}$ ranged from $0.6-1.4$ with an average value of $\sim$0.75. The resultant best fit spectral parameters variation as a function of time were considered time series with a time bin of 23.775 secs. To check for a possible time lag between the spectral parameter variation,  the  Cross-correlation function (CCF)  between the time series were computed using the HEASOFT function {\it crosscor}.  The time series were divided into 28 intervals of 16 bins, i.e. the length of each interval was 380.4 secs. Then for each pair of intervals, the cross-correlation was computed and averaged over the 28 intervals. {\it crosscor} provides the standard deviation of the averaged cross-correlation as the error. However, to further check the robustness of the observed cross-correlation to noise in the data, we implemented the pair bootstrapping simulation. We simulated 10,000 pairs of time series using the Random Subset Selection (RSS) technique \citep{pe98} and extracted the Cross-correlation function (CCF) for each pair. The 90 $\%$ confidence interval was estimated using the simulated CCF distribution at each lag.
\section{Results}
\begin{figure*}
\centering \includegraphics[width=0.47\textwidth,angle=0]{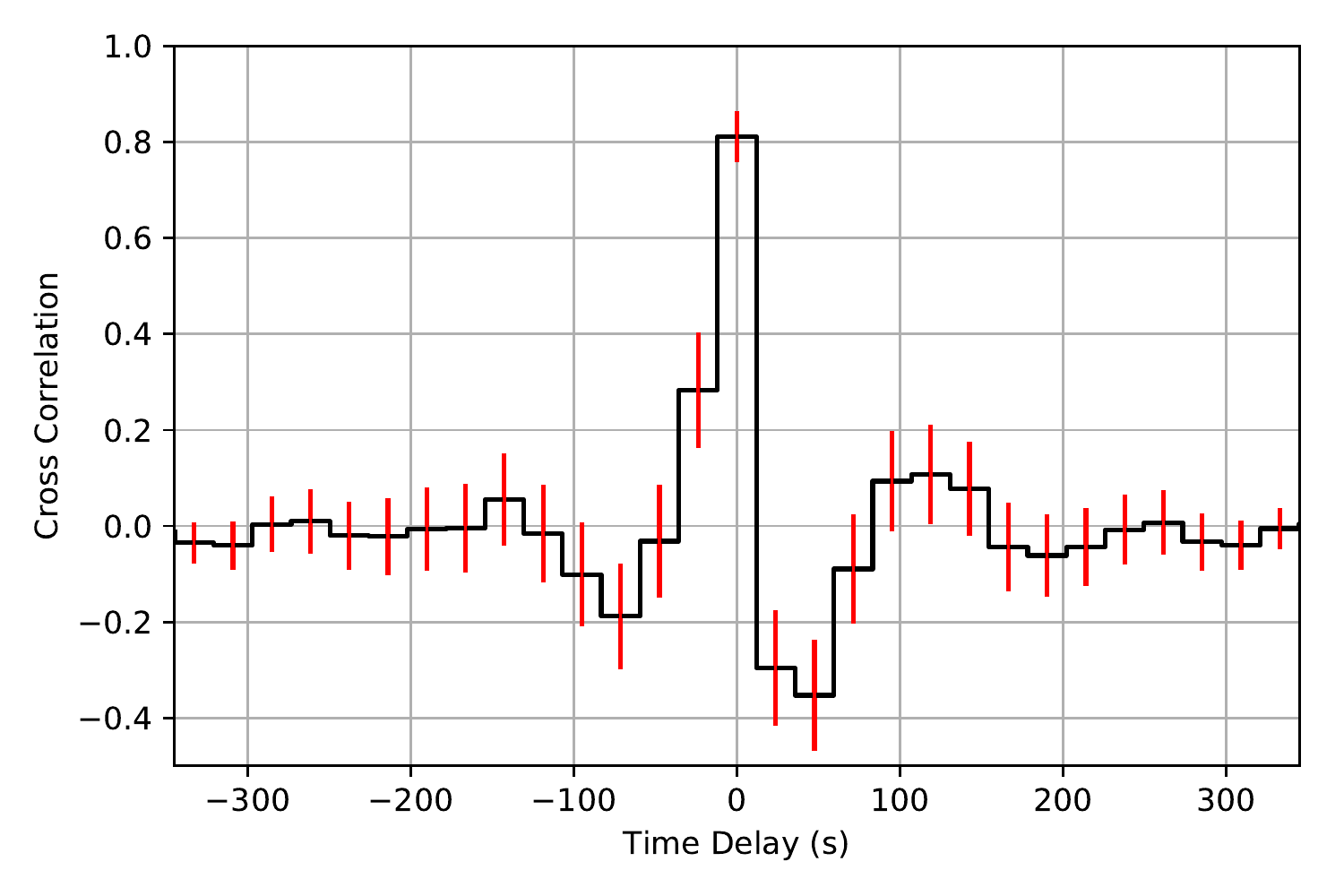}
\centering \includegraphics[width=0.47\textwidth,angle=0]{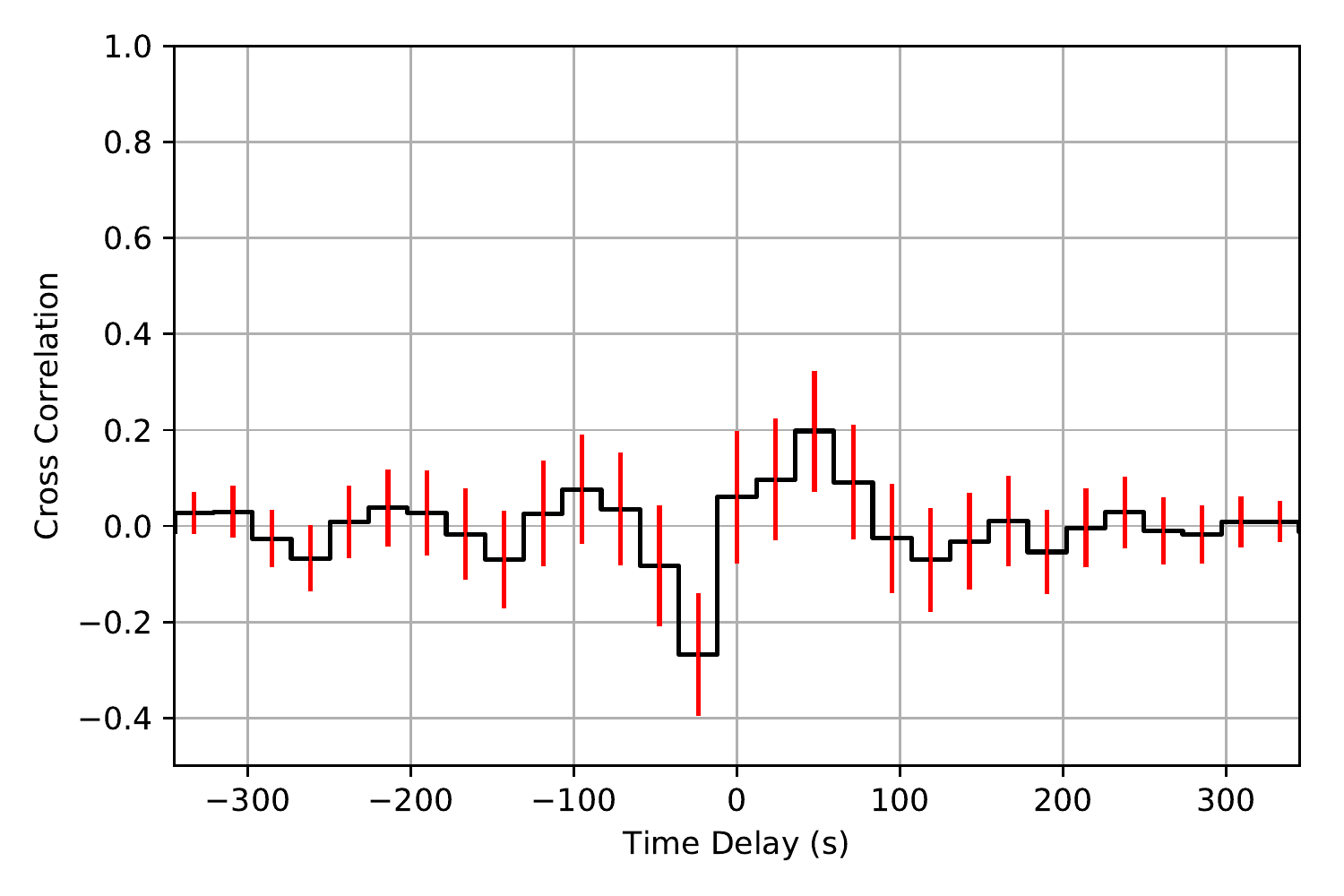}
\centering \includegraphics[width=0.47\textwidth,angle=0]{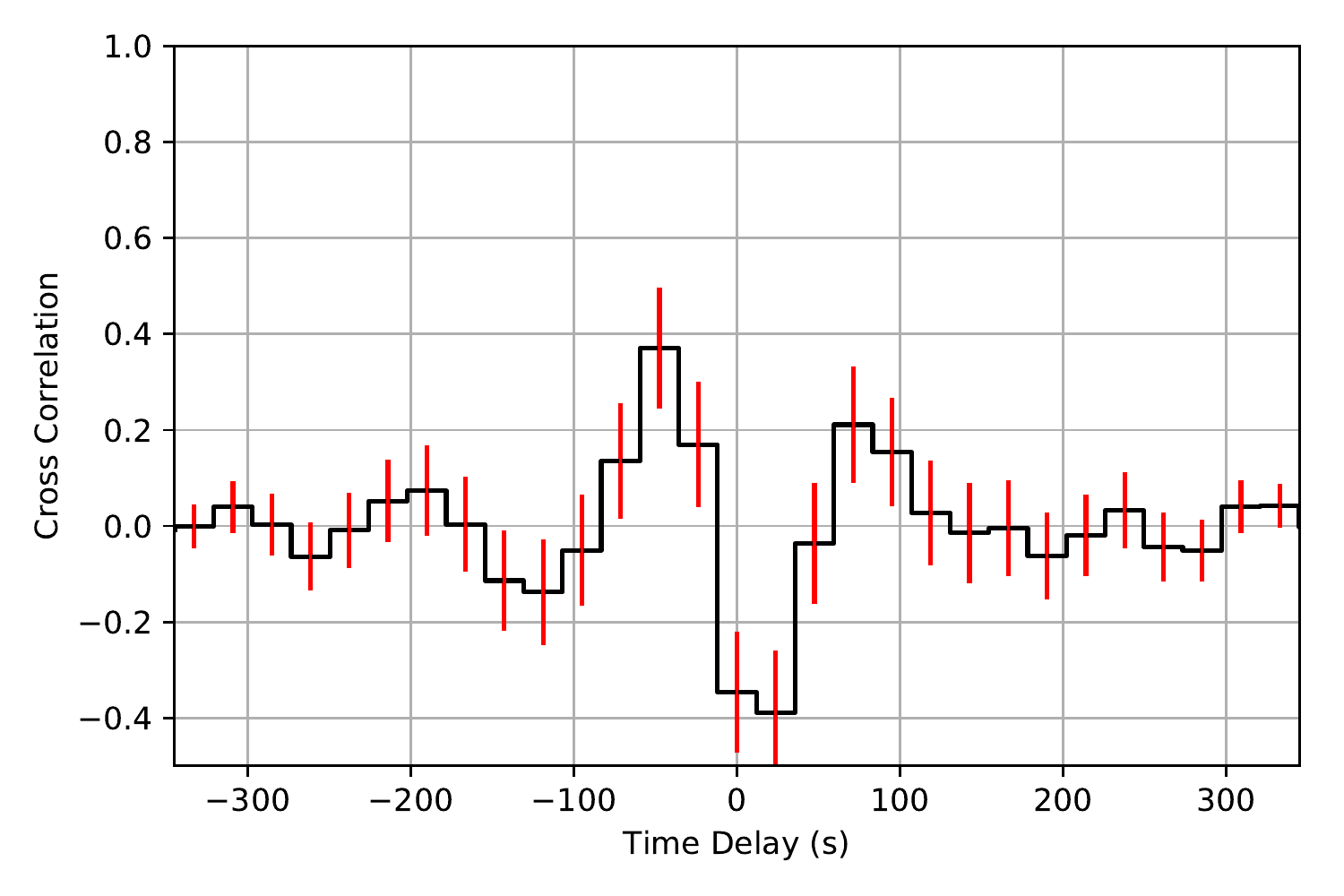}
\centering \includegraphics[width=0.47\textwidth,angle=0]{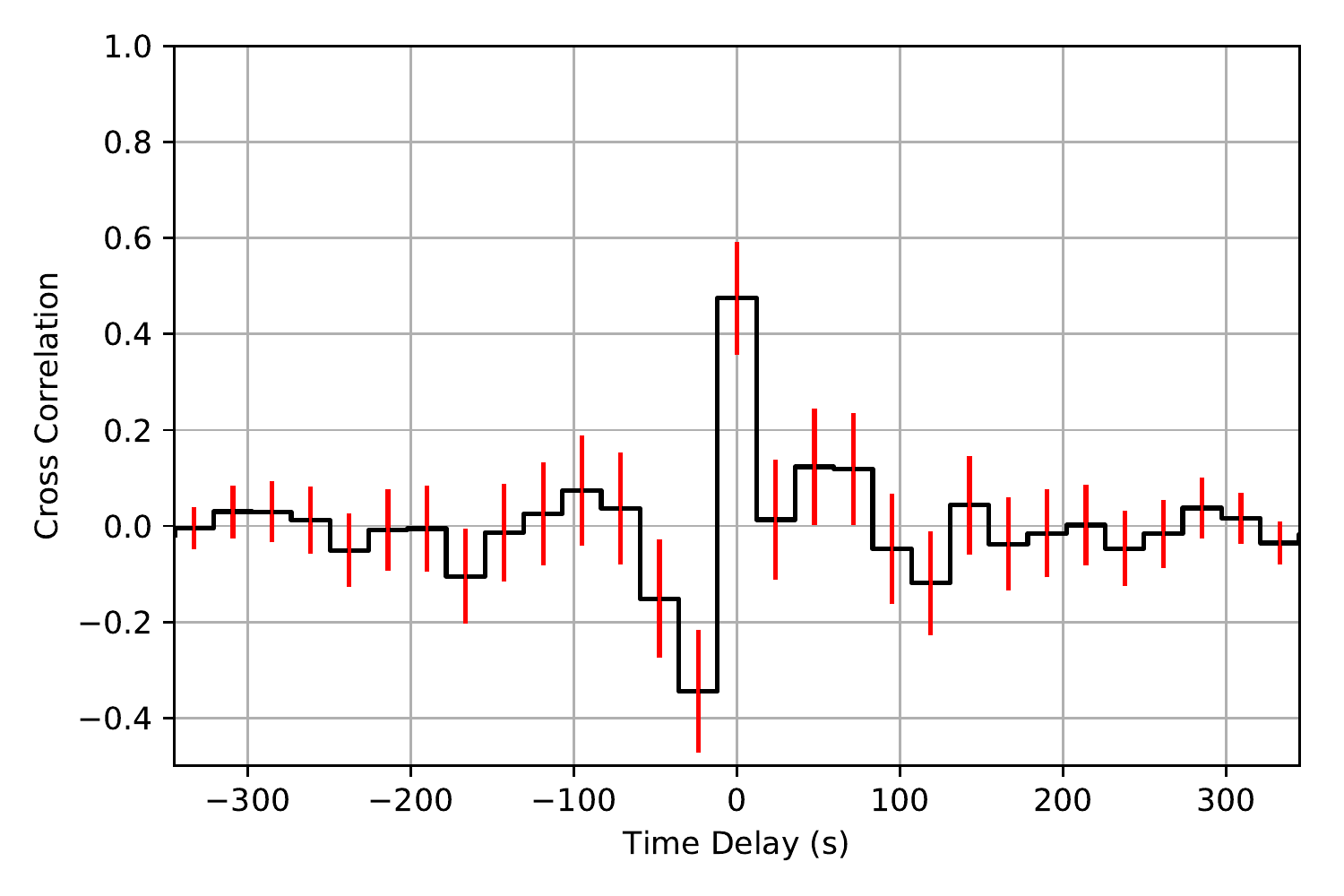}
\caption{The top left and right panel shows the Cross-Correlation functions of $r_{in}$, and $\Gamma$ with $\dot M$, respectively. The bottom left and right panel shows Cross-Correlation functions of $f_s$ with $\dot M$, and  $\Gamma$, respectively.}
\label{corr1}
\end{figure*}

The left panels of Figure \ref{parm} show the variation of the count rate and spectral parameters with time for a $\sim 1000$ sec segment of the heartbeat state. The top panel shows the count rate in a finer 2.3775 secs time-bin, while the spectral parameters are measured over 23.775 secs. The near sinusoidal variation is evident for all the parameters. The right panels of Figure \ref{parm} show the variation of the inner truncated radius, $r_{in}$, the power-law index $\Gamma$ and the scattering fraction $f_s$ with the accretion rate, $\dot M$, for the heartbeat state as well for the $\chi$ and intermediate ones. For the heartbeat state, we checked the significance of any correlation (or anti-correlation) by using a Monte Carlo simulation technique. For each data point and its associated standard deviation in the time series, we simulated data points considering a random normal distribution for the error. The Pearson correlation coefficient, $r$ and the corresponding null hypothesis probability $p$ were computed for 10,000 pairs of the simulated time-series and averaged. The correlations of $r_{in}$ and $f_s$ with $\dot M$ are highly significant ($r =0.680\pm0.001$, $p = 4.7\times 10^{-58}$  and $r =-0.361\pm0.002$, $p =6.3 \times 10^{-15}$ respectively), while for $\Gamma$ with $\dot M$ is less so ($r =0.053\pm0.003 $, $p =1.5 \times 10^{-3}$). For the $\chi$ and Intermediate states, the variation of $r_{in}$ and $f_s$ with $\dot M$ is clearly different from that of the heartbeat, while the $\Gamma$ values are in the same range. To verify whether there is any time (or phase lag) between the spectral parameter variation for the heart beat state, we computed the cross-correlation for different pairs and show them in Figure \ref{corr1}. While there are hints of possible time-lags of $\sim 100$ secs between $f_s$ and $\dot M$ and $\sim 25$ secs between $\Gamma$ and $\dot M$, the significance is not sufficient to make robust confident statements.  Note that this analysis is limited by the time-bin of the time-resolved spectroscopy of 23.775 seconds and hence does not capture the more rapid variability seen, especially in the peak of the oscillations. These faster variability  may have different spectral parameter correlations than that reported here.


\section{Discussion}     
Our observations reveal correlated variation of spectral parameters in the heartbeat state. We find that the total flux, accretion rate $\dot M$, inner radius $r_{in}$ show periodic variation, nearly in phase with one another.  Here we discuss possible physical mechanism which may explain these observations.  
\subsection{Viscous time-scale}
The solution for the standard accretion disc model has been derived by \citet{sh73} and  later relativistic corrections were included by  \citet{no73}. In the radiation pressure dominated region where the opacity is mainly due to electron-electron scattering, we have calculated the viscous time-scale for heartbeat state using equation (5.9.10) from \citet{no73}. 
For $\dot M = 0.72\pm{0.02}$ $\times$10$^{18}$ gm/s and $\alpha$=0.1 the material falling from the outer part of the disc $\sim$20 R$_g$ takes about 100-150 secs to reach the innermost stable circular orbit (ISCO). We have used a high value of the dimensionless spin parameter $a=0.98$ as found by \citep{mc06,bl09,mi20,sr20,ho20} for the viscous time-scale calculation.
\begin{figure}
\centering \includegraphics[width=0.47\textwidth,angle=0]{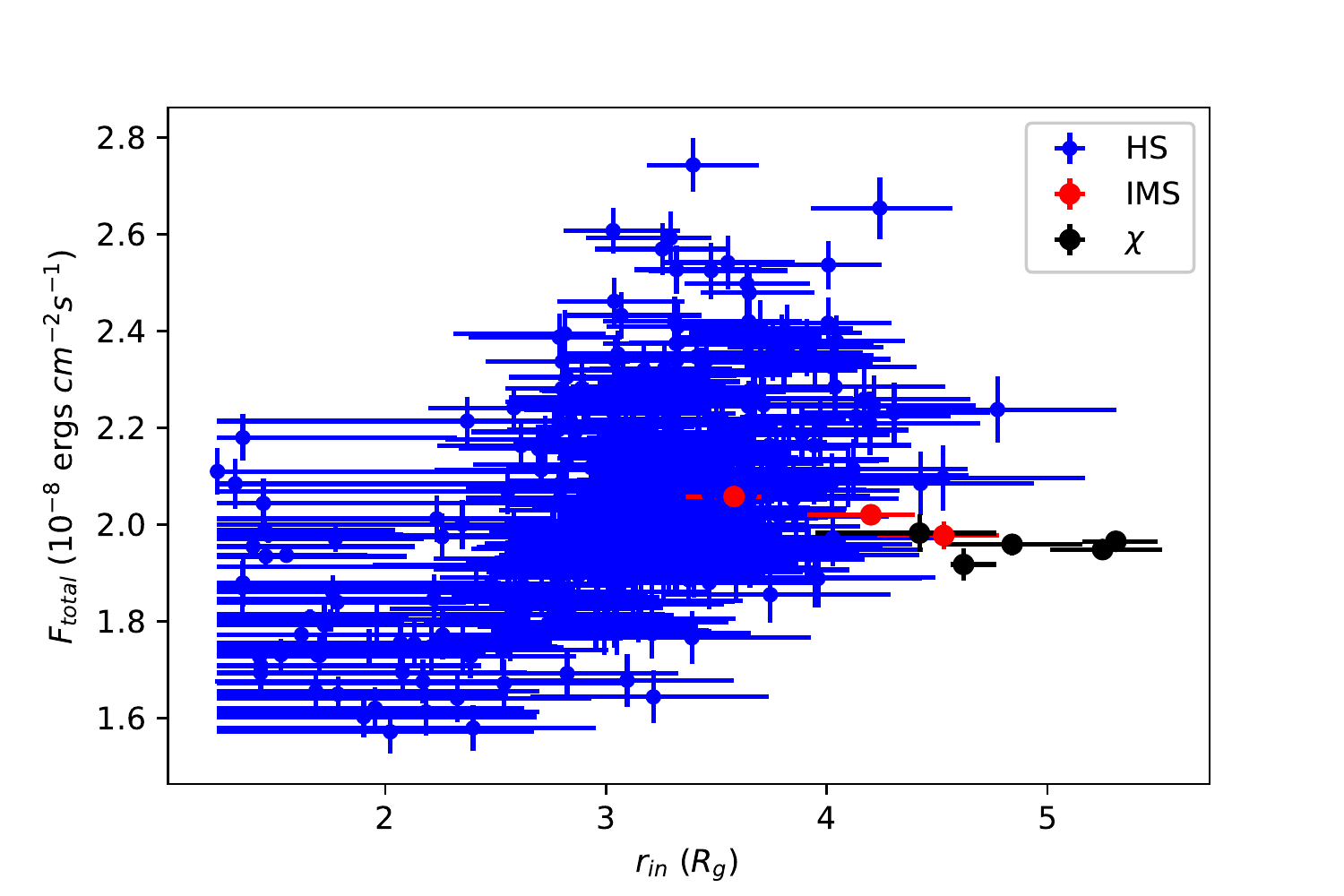}
\caption{The variation of the total flux with inner disc radii for $\chi$ class, IMS and HS.}
\label{accretion_rate_flux}
\end{figure} 
\begin{figure*}
\centering \includegraphics[scale=0.55]{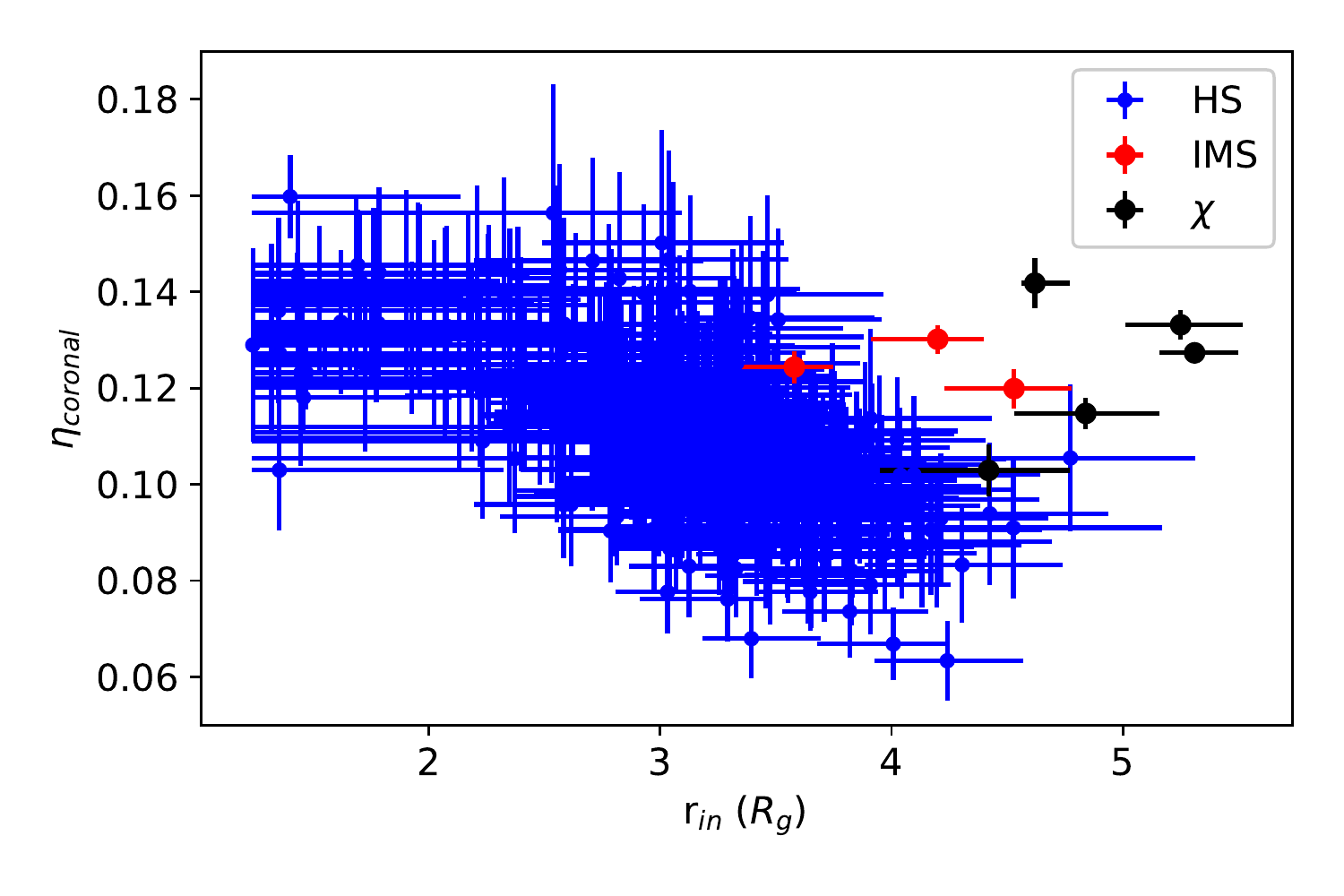}
\centering \includegraphics[scale=0.55]{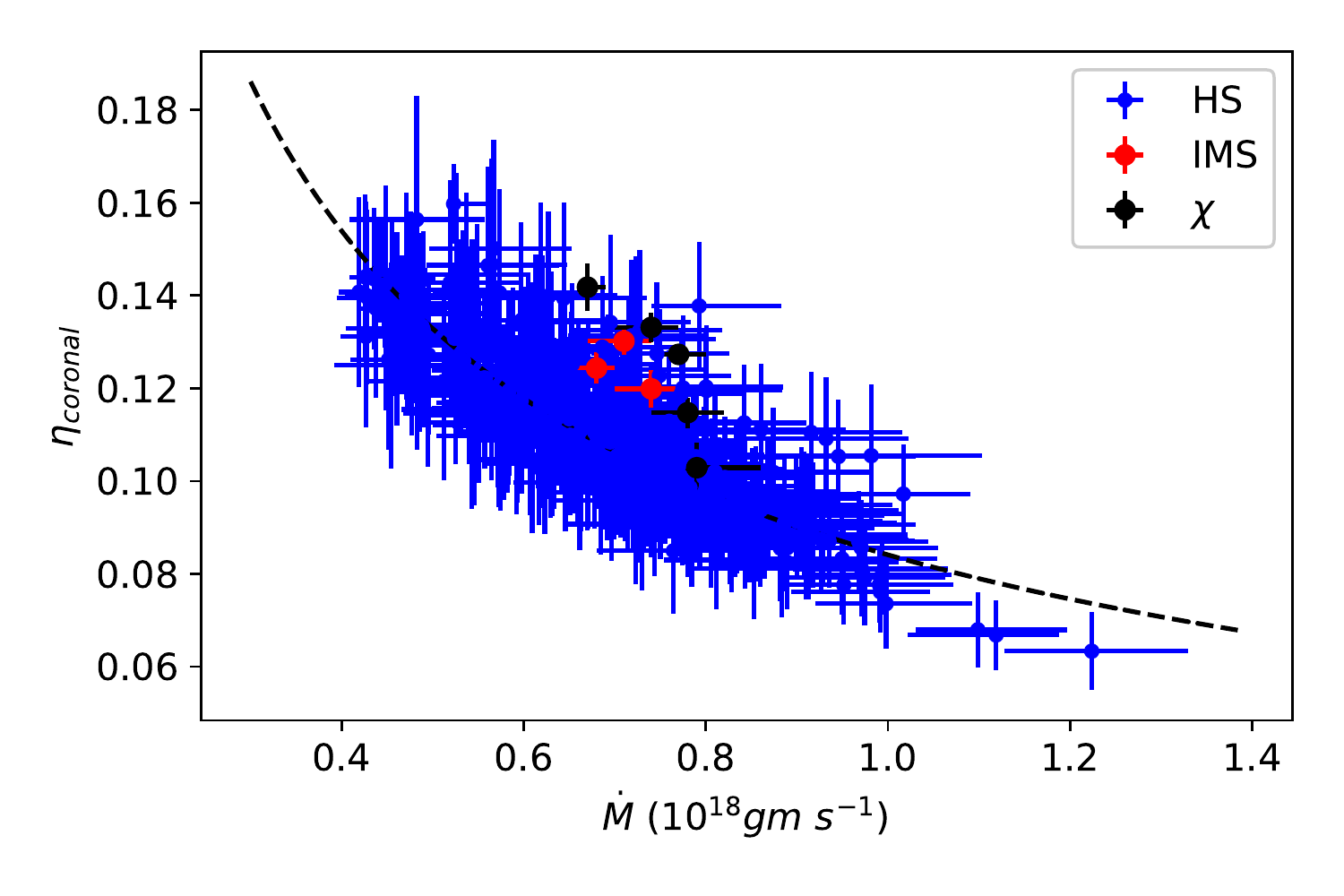}
\caption{The variation of coronal efficiency ($\eta _{coronal}$) for $\chi$ class, IMS and HS with inner disc radii and accretion rate are shown in left and right panel respectively. The dotted line in the right panel represents the best fit curve $\eta_{coronal} \propto {\dot M}^{-b}$, with  $b = 0.66 \pm 0.02$}
\label{figure12}
\end{figure*}
\subsection{Oscillation in $\dot M$ as response of inward propagation of fluctuation}
As the source undergoes transitions from $\chi$ class to Heartbeat class via IMS, $r_{in}$ starts to oscillate in response to change in accretion rate, as shown in the figure \ref{parm}. The oscillatory nature of $\dot M$ has been also observed by \citet{mi03} in the hard-state of GRS 1915+105 and they have associated the variation in the local accretion rate to the loss of matter in the inner region of the disc due to instability. \citet{ne12} argued that all the heartbeat state variability could be explained as a result of the combination of thermal-viscous radiation pressure instability \citep{li74,ta97,na00}, a local Eddington limit, and sudden radiation-pressure driven evaporation of inner disc \citep{fu04,he07,li09}. In our observations, we have observed a correlation between total unabsorbed 0.1-80.0 keV X-ray flux and inner disc radii for full HS oscillation (shown in figure \ref{accretion_rate_flux}).  This is consistent with \citet{ni11,ne12} , who also observe a steady increase in the disc radii during the phase in which the flux shows a slow rise. This is followed by a sharp fall in radii over a very short time interval as luminosity approaches its peak value. In our analysis, we have not tested the behaviour at such short time intervals. We point out that the luminosity in \citet{ni11,ne12}  was close to the Eddington limit and in our case it is approximately $0.1L_{edd}$. Furthermore,  our oscillation period is twice as large as their period.
\subsection{Correlation between coronal efficiency and accretion rate}
The heating rate of the corona should be equal to excess luminosity emitted by it, $L_{ac}$ which would be the  difference in the luminosity of the corona $L_c$ and the luminosity of the seed photons entering it. Hence $L_{ac} = L_c - f_sL_d$, where $f_s$ is the fraction of the disc luminosity $L_d$ that enters the corona. We can then define a coronal radiative efficiency as $\eta_{coronal} \equiv L_{ac}/\dot M c^2$. In the spectral model used, the total unabsorbed flux $F_{tot} = (1-f_s)F_d + F_c$, where $F_d$ and $F_c$ are the disc and coronal fluxes. For isotropic emission, the corona luminosity  $L_c = 4 \pi D^2 F_c $, where $D$ is the distance to the source. The disc luminosity  $L_d = 2 \pi D^2 F_d/cos(i) \sim 4 \pi D^2 F_d$ for an inclination angle $i \sim 60 ^{\circ}$. From this we can compute the excess coronal luminosity as $L_{ac} = 4 \pi D^2 (F_{tot}-F_d)$. We estimate the total unabsorbed flux and the disc flux using the {\sc{Xspec}} model ``cflux". 
For a distance of $8.5$ kpc and using the best fit values of the fluxes and the accretion
rate $\dot M$  for each spectral fitting, we estimate the coronal efficiency and study its correlation with disc accretion rate and inner disc radii.


Figure \ref{figure12} shows the variation of coronal efficiency $\eta_{coronal}$ with inner disc radii and accretion rate. The variation of the coronal efficiency with accretion rate is relatively similar for all the states (Right panel of Figure \ref{figure12}). For the same accretion rate, the $\chi$ and IMS states seem to have slightly higher efficiencies as compared to the HS state, but it is within the dispersion
seen in the HS state. The dependence can be represented as $\eta_{coronal} \propto \dot M^{-2/3}$ shown as a dashed line in the Figure \ref{figure12}. For the HS state, time resolved spectral results show that $\eta_{coronal}$ also  varies inversely with inner radii (left panel of Figure \ref{figure12}).  However, for the $\chi$ and IMS states, while the efficiencies have similar values as some of the HS segments, the inner disc radii are significantly different. In other words, the system can exhibit nearly the same coronal efficiency for different values of the inner radius depending on the state.

 
The coronal emission may arise from a hot radiatively inefficient flow at radii less than the inner radius of the standard disc. However, for an advection-dominated accretion flow (ADAF) the efficiency is proportional to the accretion rate \citep[e.g.][]{na94}. While a pure self-similar ADAF solution may not be strictly valid when the standard disc is truncated at small radii, the measured efficiency being $\propto \dot M^{-2/3}$ argues against the emission being from an inner hot flow. Alternatively, the coronal emission could be from a corona located above and below the standard disc (sandwich geometry), where a fraction of the gravitational energy released is dissipated in the corona \citep[e.g.][]{ha93}. For such a geometry the coronal efficiency should decrease with inner disc radii and indeed for the HS observations this is what is see in Figure \ref{figure12}. Thus, for the HS state the results seem to indicate that the hard X-ray emission arises from a corona above and below the disc and the inner regions at radii less then inner disc radius does not have a significant contribution.\\
For the $\chi$ and Intermediate states, the coronal efficiency is similar to the HS but the inner radii are larger. This could imply that basic geometry of the system is different for these states and perhaps here the hard X-ray emission arises from the inner hot flow. If one insists and assumes that the geometry of all the states are qualitatively same, then the results imply that the coronal efficiency is  $\propto \dot M^{-2/3}$ generically and the efficiency does not depend on the inner radii and the variation with inner radii seen for the HS would then be an indirect one. In such an assumption, neither a hot inner disc nor a corona on top and bottom of the disc would be a favoured and a speculative interpretation could be if the corona is being powered by the spin of the black hole i.e. by the Blandford-Znajek (BZ) mechanism \citep{bl77}. The BZ mechanism has typically been invoked to explain the powerful jets seen in black hole systems including GRS 1915+105. It has been proposed that the base of these jets is the corona which is the source of hard X-ray emission \citep{ma01}. Note that even if the energy source of the corona is the same as that of the jet (i.e. the BZ mechanism), the dominant radiative process for the corona could still be thermal Comptonization instead of synchrotron by which the larger scale jet radiates. In the BZ mechanism the energy is extracted from the spin of the black hole and the extraction rate is proportional to $B_\phi^2 $, where $B_\phi^2 $ is the poloidal magnetic field treading the black hole. If we assume that $B_\phi^2 $ is proportional to the cube root of the accretion rate falling onto the black hole and the radiative efficiency is independent, then the coronal flux would be $F_{coronal} \propto \dot M^{1/3}$ (i.e. $\eta_{coronal} \propto \dot M^{-2/3}$) as reported here. Moreover, since the process may depend only on the amount of matter falling into the black hole, it may be independent of the inner radius of the standard disc. The details of the BZ mechanism are complex and uncertain and in particular, the dependence of $B_{\phi}$ on the accretion rate is theoretically not well understood.
Hence a quantitative study as to whether the observed correlation is consistent with the mechanism is beyond the scope of this paper. However, for the states considered here the source does not exhibit powerful jets and hence it is not clear that the BZ mechanism would be active.\\
A similar analysis for the other states of GRS
1915+105 (and other black hole systems) need to be undertaken to ascertain the scope and universality of the coronal
efficiency and accretion rate correlation, which in turn has the
high possibility of distinguishing the different mechanisms by
which the corona is heated.
\section{Summary and Conclusions}
\begin{itemize}
\item We have performed time-resolved spectroscopy for the heartbeat state of GRS 1915+105 and found that the heartbeat state oscillation could be described as coordinated variation of accretion rate, inner disc radii and comptonised flux occurring at 0.1$L_{edd}$. 
\item A correlation between coronal efficiency and accretion rate is found for $\chi$, IMS and heartbeat state which is approximately $\eta_{coronal} \propto \dot M^{-2/3}$. For the HS state the efficiency decreases with inner disc radii. However, for the IMS and $\chi$ states the efficiency is similar to that of the HS despite the inner radii being larger.

\end{itemize}
\section*{Acknowledgements}
We thank the referee for constructive suggestions, which
were very helpful in improving the article. We thank the members of LAXPC $\&$ SXT instrument team for their contribution to the development of the instrument. We also acknowledge the contributions of the AstroSat project team at ISAC. This research work makes use of data from the AstroSat mission of the Indian Space Research Organisation (ISRO), archived at the Indian Space Science Data Centre (ISSDC). DR acknowledges Samuzal Barua, and Vaishak Prasad for enlightening discussions.
\section*{Data Availability}
The datasets were derived from sources in the public domain: \href{https://webapps.issdc.gov.in/astro_archive/archive/Home.jsp}{AstroSat Archive}. The software and tools used to process the data are available at \href{http://astrosat-ssc.iucaa.in/?q=laxpcData}{LAXPC Software}, \href{http://astrosat-ssc.iucaa.in/?q=sxtData}{SXT Software}.
\def\aj{AJ} \def\actaa{Acta Astron.}  \def\araa{ARA\&A} \def\apj{ApJ}
\def\apjl{ApJ} \def\apjs{ApJS} \def\ao{Appl.~Opt.}  \def\apss{Ap\&SS}
\def\aap{A\&A} \def\aapr{A\&A~Rev.}  \def\aaps{A\&AS} \def\azh{AZh}
\def\baas{BAAS} \def\bac{Bull. astr. Inst. Czechosl.}
\def\caa{Chinese Astron. Astrophys.}  \def\cjaa{Chinese
  J. Astron. Astrophys.}  \def\icarus{Icarus} \def\jcap{J. Cosmology
  Astropart. Phys.}  \def\jrasc{JRASC} \def\mnras{MNRAS}
\def\memras{MmRAS} \def\na{New A} \def\nar{New A Rev.}
\def\pasa{PASA} \def\pra{Phys.~Rev.~A} \def\prb{Phys.~Rev.~B}
\def\prc{Phys.~Rev.~C} \def\prd{Phys.~Rev.~D} \def\pre{Phys.~Rev.~E}
\def\prl{Phys.~Rev.~Lett.}  \def\pasp{PASP} \def\pasj{PASJ}
\def\qjras{QJRAS} \def\rmxaa{Rev. Mexicana Astron. Astrofis.}
\def\skytel{S\&T} \def\solphys{Sol.~Phys.}  \def\sovast{Soviet~Ast.}
\def\ssr{Space~Sci.~Rev.}  \def\zap{ZAp} \def\nat{Nature}
\def\iaucirc{IAU~Circ.}  \def\aplett{Astrophys.~Lett.}
\def\apspr{Astrophys.~Space~Phys.~Res.}
\def\bain{Bull.~Astron.~Inst.~Netherlands}
\def\fcp{Fund.~Cosmic~Phys.}  \def\gca{Geochim.~Cosmochim.~Acta}
\def\grl{Geophys.~Res.~Lett.}  \def\jcp{J.~Chem.~Phys.}
\def\jgr{J.~Geophys.~Res.}
\def\jqsrt{J.~Quant.~Spec.~Radiat.~Transf.}
\def\memsai{Mem.~Soc.~Astron.~Italiana} \def\nphysa{Nucl.~Phys.~A}
\def\physrep{Phys.~Rep.}  \def\physscr{Phys.~Scr}
\def\planss{Planet.~Space~Sci.}  \def\procspie{Proc.~SPIE}
\let\astap=\aap \let\apjlett=\apjl \let\apjsupp=\apjs \let\applopt=\ao

\bibliographystyle{mnras}
\bibliography{bibliography} 

\begin{thebibliography}{}
\makeatletter
\relax
\def\mn@urlcharsother{\let\do\@makeother \do\$\do\&\do\#\do\^\do\_\do\%\do\~}
\def\mn@doi{\begingroup\mn@urlcharsother \@ifnextchar [ {\mn@doi@}
  {\mn@doi@[]}}
\def\mn@doi@[#1]#2{\def\@tempa{#1}\ifx\@tempa\@empty \href
  {http://dx.doi.org/#2} {doi:#2}\else \href {http://dx.doi.org/#2} {#1}\fi
  \endgroup}
\def\mn@eprint#1#2{\mn@eprint@#1:#2::\@nil}
\def\mn@eprint@arXiv#1{\href {http://arxiv.org/abs/#1} {{\tt arXiv:#1}}}
\def\mn@eprint@dblp#1{\href {http://dblp.uni-trier.de/rec/bibtex/#1.xml}
  {dblp:#1}}
\def\mn@eprint@#1:#2:#3:#4\@nil{\def\@tempa {#1}\def\@tempb {#2}\def\@tempc
  {#3}\ifx \@tempc \@empty \let \@tempc \@tempb \let \@tempb \@tempa \fi \ifx
  \@tempb \@empty \def\@tempb {arXiv}\fi \@ifundefined
  {mn@eprint@\@tempb}{\@tempb:\@tempc}{\expandafter \expandafter \csname
  mn@eprint@\@tempb\endcsname \expandafter{\@tempc}}}

\bibitem[\protect\citeauthoryear{{Agrawal}}{{Agrawal}}{2006}]{ag06}
{Agrawal} P.~C.,  2006, \mn@doi [Advances in Space Research]
  {10.1016/j.asr.2006.03.038}, \href
  {https://ui.adsabs.harvard.edu/abs/2006AdSpR..38.2989A} {38, 2989}

\bibitem[\protect\citeauthoryear{{Antia} et~al.,}{{Antia} et~al.}{2017}]{An17}
{Antia} H.~M.,  et~al., 2017, \mn@doi [\apjs] {10.3847/1538-4365/aa7a0e}, \href
  {http://adsabs.harvard.edu/abs/2017ApJS..231...10A} {231, 10}

\bibitem[\protect\citeauthoryear{{Belloni}, {Klein-Wolt}, {M{\'e}ndez}, {van
  der Klis}  \& {van Paradijs}}{{Belloni} et~al.}{2000}]{be00}
{Belloni} T.,  {Klein-Wolt} M.,  {M{\'e}ndez} M.,  {van der Klis} M.,   {van
  Paradijs} J.,  2000, \aap, \href
  {http://adsabs.harvard.edu/abs/2000A%26A...355..271B} {355, 271}

\bibitem[\protect\citeauthoryear{{Blandford} \& {Znajek}}{{Blandford} \&
  {Znajek}}{1977}]{bl77}
{Blandford} R.~D.,  {Znajek} R.~L.,  1977, \mn@doi [\mnras]
  {10.1093/mnras/179.3.433}, \href
  {https://ui.adsabs.harvard.edu/abs/1977MNRAS.179..433B} {179, 433}

\bibitem[\protect\citeauthoryear{{Blum}, {Miller}, {Fabian}, {Miller}, {Homan},
  {van der Klis}, {Cackett}  \& {Reis}}{{Blum} et~al.}{2009}]{bl09}
{Blum} J.~L.,  {Miller} J.~M.,  {Fabian} A.~C.,  {Miller} M.~C.,  {Homan} J.,
  {van der Klis} M.,  {Cackett} E.~M.,   {Reis} R.~C.,  2009, \mn@doi [\apj]
  {10.1088/0004-637X/706/1/60}, \href
  {https://ui.adsabs.harvard.edu/abs/2009ApJ...706...60B} {706, 60}

\bibitem[\protect\citeauthoryear{{Fukue}}{{Fukue}}{2004}]{fu04}
{Fukue} J.,  2004, \mn@doi [\pasj] {10.1093/pasj/56.3.569}, \href
  {https://ui.adsabs.harvard.edu/abs/2004PASJ...56..569F} {56, 569}

\bibitem[\protect\citeauthoryear{{Greiner}, {Morgan}  \& {Remillard}}{{Greiner}
  et~al.}{1996}]{gr96}
{Greiner} J.,  {Morgan} E.~H.,   {Remillard} R.~A.,  1996, \mn@doi [\apjl]
  {10.1086/310402}, \href {http://adsabs.harvard.edu/abs/1996ApJ...473L.107G}
  {473, L107}

\bibitem[\protect\citeauthoryear{{Greiner}, {Cuby}, {McCaughrean},
  {Castro-Tirado}  \& {Mennickent}}{{Greiner} et~al.}{2001}]{gr01}
{Greiner} J.,  {Cuby} J.~G.,  {McCaughrean} M.~J.,  {Castro-Tirado} A.~J.,
  {Mennickent} R.~E.,  2001, \mn@doi [\aap] {10.1051/0004-6361:20010771}, \href
  {http://adsabs.harvard.edu/abs/2001A%26A...373L..37G} {373, L37}

\bibitem[\protect\citeauthoryear{{Haardt} \& {Maraschi}}{{Haardt} \&
  {Maraschi}}{1993}]{ha93}
{Haardt} F.,  {Maraschi} L.,  1993, \mn@doi [\apj] {10.1086/173020}, \href
  {https://ui.adsabs.harvard.edu/abs/1993ApJ...413..507H} {413, 507}

\bibitem[\protect\citeauthoryear{{Hannikainen} et~al.,}{{Hannikainen}
  et~al.}{2003}]{ha03}
{Hannikainen} D.~C.,  et~al., 2003, \mn@doi [\aap]
  {10.1051/0004-6361:20031444}, \href
  {http://adsabs.harvard.edu/abs/2003A%26A...411L.415H} {411, L415}

\bibitem[\protect\citeauthoryear{{Heinzeller} \& {Duschl}}{{Heinzeller} \&
  {Duschl}}{2007}]{he07}
{Heinzeller} D.,  {Duschl} W.~J.,  2007, \mn@doi [\mnras]
  {10.1111/j.1365-2966.2006.11233.x}, \href
  {https://ui.adsabs.harvard.edu/abs/2007MNRAS.374.1146H} {374, 1146}

\bibitem[\protect\citeauthoryear{{Klein-Wolt}, {Fender}, {Pooley}, {Belloni},
  {Migliari}, {Morgan}  \& {van der Klis}}{{Klein-Wolt} et~al.}{2002}]{ke02}
{Klein-Wolt} M.,  {Fender} R.~P.,  {Pooley} G.~G.,  {Belloni} T.,  {Migliari}
  S.,  {Morgan} E.~H.,   {van der Klis} M.,  2002, \mn@doi [\mnras]
  {10.1046/j.1365-8711.2002.05223.x}, \href
  {http://adsabs.harvard.edu/abs/2002MNRAS.331..745K} {331, 745}

\bibitem[\protect\citeauthoryear{{Lightman} \& {Eardley}}{{Lightman} \&
  {Eardley}}{1974}]{li74}
{Lightman} A.~P.,  {Eardley} D.~M.,  1974, \mn@doi [\apjl] {10.1086/181377},
  \href {https://ui.adsabs.harvard.edu/abs/1974ApJ...187L...1L} {187, L1}

\bibitem[\protect\citeauthoryear{{Lin}, {Remillard}  \& {Homan}}{{Lin}
  et~al.}{2009}]{li09}
{Lin} D.,  {Remillard} R.~A.,   {Homan} J.,  2009, in American Astronomical
  Society Meeting Abstracts \#213. p. 603.03

\bibitem[\protect\citeauthoryear{{Liu}, {Ji}, {Bambi}, {Jain}, {Misra},
  {Rawat}, {Yadav}  \& {Zhang}}{{Liu} et~al.}{2020}]{ho20}
{Liu} H.,  {Ji} L.,  {Bambi} C.,  {Jain} P.,  {Misra} R.,  {Rawat} D.,  {Yadav}
  J.~S.,   {Zhang} Y.,  2020, arXiv e-prints, \href
  {https://ui.adsabs.harvard.edu/abs/2020arXiv201201825L} {p. arXiv:2012.01825}

\bibitem[\protect\citeauthoryear{{Markoff}, {Falcke}  \& {Fender}}{{Markoff}
  et~al.}{2001}]{ma01}
{Markoff} S.,  {Falcke} H.,   {Fender} R.,  2001, \mn@doi [\aap]
  {10.1051/0004-6361:20010420}, \href
  {https://ui.adsabs.harvard.edu/abs/2001A&A...372L..25M} {372, L25}

\bibitem[\protect\citeauthoryear{{McClintock}, {Shafee}, {Narayan},
  {Remillard}, {Davis}  \& {Li}}{{McClintock} et~al.}{2006}]{mc06}
{McClintock} J.~E.,  {Shafee} R.,  {Narayan} R.,  {Remillard} R.~A.,  {Davis}
  S.~W.,   {Li} L.-X.,  2006, \mn@doi [\apj] {10.1086/508457}, \href
  {http://adsabs.harvard.edu/abs/2006ApJ...652..518M} {652, 518}

\bibitem[\protect\citeauthoryear{{Migliari} \& {Belloni}}{{Migliari} \&
  {Belloni}}{2003}]{mi03}
{Migliari} S.,  {Belloni} T.,  2003, \mn@doi [\aap]
  {10.1051/0004-6361:20030484}, \href
  {http://adsabs.harvard.edu/abs/2003A%26A...404..283M} {404, 283}

\bibitem[\protect\citeauthoryear{{Misra}, {Rawat}, {Yadav}  \& {Jain}}{{Misra}
  et~al.}{2020}]{mi20}
{Misra} R.,  {Rawat} D.,  {Yadav} J.~S.,   {Jain} P.,  2020, \mn@doi [\apjl]
  {10.3847/2041-8213/ab6ddc}, \href
  {https://ui.adsabs.harvard.edu/abs/2020ApJ...889L..36M} {889, L36}

\bibitem[\protect\citeauthoryear{{Mu{\~n}oz-Darias}, {Fender}, {Motta}  \&
  {Belloni}}{{Mu{\~n}oz-Darias} et~al.}{2014}]{da14}
{Mu{\~n}oz-Darias} T.,  {Fender} R.~P.,  {Motta} S.~E.,   {Belloni} T.~M.,
  2014, \mn@doi [\mnras] {10.1093/mnras/stu1334}, \href
  {http://adsabs.harvard.edu/abs/2014MNRAS.443.3270M} {443, 3270}

\bibitem[\protect\citeauthoryear{{Narayan} \& {Yi}}{{Narayan} \&
  {Yi}}{1994}]{na94}
{Narayan} R.,  {Yi} I.,  1994, \mn@doi [\apjl] {10.1086/187381}, \href
  {https://ui.adsabs.harvard.edu/abs/1994ApJ...428L..13N} {428, L13}

\bibitem[\protect\citeauthoryear{{Nayakshin}, {Rappaport}  \&
  {Melia}}{{Nayakshin} et~al.}{2000}]{na00}
{Nayakshin} S.,  {Rappaport} S.,   {Melia} F.,  2000, \mn@doi [\apj]
  {10.1086/308860}, \href
  {https://ui.adsabs.harvard.edu/abs/2000ApJ...535..798N} {535, 798}

\bibitem[\protect\citeauthoryear{{Neilsen}, {Remillard}  \& {Lee}}{{Neilsen}
  et~al.}{2011}]{ni11}
{Neilsen} J.,  {Remillard} R.~A.,   {Lee} J.~C.,  2011, \mn@doi [\apj]
  {10.1088/0004-637X/737/2/69}, \href
  {https://ui.adsabs.harvard.edu/abs/2011ApJ...737...69N} {737, 69}

\bibitem[\protect\citeauthoryear{{Neilsen}, {Remillard}  \& {Lee}}{{Neilsen}
  et~al.}{2012}]{ne12}
{Neilsen} J.,  {Remillard} R.~A.,   {Lee} J.~C.,  2012, \mn@doi [\apj]
  {10.1088/0004-637X/750/1/71}, \href
  {http://adsabs.harvard.edu/abs/2012ApJ...750...71N} {750, 71}

\bibitem[\protect\citeauthoryear{{Neilsen}, {Homan}, {Steiner}, {Marcel},
  {Cackett}, {Remillard}  \& {Gendreau}}{{Neilsen} et~al.}{2020}]{ne20}
{Neilsen} J.,  {Homan} J.,  {Steiner} J.~F.,  {Marcel} G.,  {Cackett} E.,
  {Remillard} R.~A.,   {Gendreau} K.,  2020, \mn@doi [\apj]
  {10.3847/1538-4357/abb598}, \href
  {https://ui.adsabs.harvard.edu/abs/2020ApJ...902..152N} {902, 152}

\bibitem[\protect\citeauthoryear{{Novikov} \& {Thorne}}{{Novikov} \&
  {Thorne}}{1973}]{no73}
{Novikov} I.~D.,  {Thorne} K.~S.,  1973, in Black Holes (Les Astres Occlus). pp
  343--450

\bibitem[\protect\citeauthoryear{{Pahari}, {Yadav}  \&
  {Bhattacharyya}}{{Pahari} et~al.}{2014}]{pa14}
{Pahari} M.,  {Yadav} J.~S.,   {Bhattacharyya} S.,  2014, \mn@doi [\apj]
  {10.1088/0004-637X/783/2/141}, \href
  {https://ui.adsabs.harvard.edu/abs/2014ApJ...783..141P} {783, 141}

\bibitem[\protect\citeauthoryear{{Paul}, {Agrawal}, {Rao}, {Vahia}, {Yadav},
  {Seetha}  \& {Kasturirangan}}{{Paul} et~al.}{1998}]{pa98}
{Paul} B.,  {Agrawal} P.~C.,  {Rao} A.~R.,  {Vahia} M.~N.,  {Yadav} J.~S.,
  {Seetha} S.,   {Kasturirangan} K.,  1998, \mn@doi [\apjl] {10.1086/311087},
  \href {https://ui.adsabs.harvard.edu/abs/1998ApJ...492L..63P} {492, L63}

\bibitem[\protect\citeauthoryear{{Peterson}, {Wanders}, {Horne}, {Collier},
  {Alexander}, {Kaspi}  \& {Maoz}}{{Peterson} et~al.}{1998}]{pe98}
{Peterson} B.~M.,  {Wanders} I.,  {Horne} K.,  {Collier} S.,  {Alexander} T.,
  {Kaspi} S.,   {Maoz} D.,  1998, \mn@doi [\pasp] {10.1086/316177}, \href
  {https://ui.adsabs.harvard.edu/abs/1998PASP..110..660P} {110, 660}

\bibitem[\protect\citeauthoryear{{Rawat} et~al.,}{{Rawat} et~al.}{2019}]{ra19}
{Rawat} D.,  et~al., 2019, \mn@doi [\apj] {10.3847/1538-4357/aaefed}, \href
  {http://adsabs.harvard.edu/abs/2019ApJ...870....4R} {870, 4}

\bibitem[\protect\citeauthoryear{{Reid}, {McClintock}, {Steiner}, {Steeghs},
  {Remillard}, {Dhawan}  \& {Narayan}}{{Reid} et~al.}{2014}]{re14}
{Reid} M.~J.,  {McClintock} J.~E.,  {Steiner} J.~F.,  {Steeghs} D.,
  {Remillard} R.~A.,  {Dhawan} V.,   {Narayan} R.,  2014, \mn@doi [\apj]
  {10.1088/0004-637X/796/1/2}, \href
  {http://adsabs.harvard.edu/abs/2014ApJ...796....2R} {796, 2}

\bibitem[\protect\citeauthoryear{{Shakura} \& {Sunyaev}}{{Shakura} \&
  {Sunyaev}}{1973}]{sh73}
{Shakura} N.~I.,  {Sunyaev} R.~A.,  1973, \aap, \href
  {http://adsabs.harvard.edu/abs/1973A%26A....24..337S} {24, 337}

\bibitem[\protect\citeauthoryear{{Singh} et~al.,}{{Singh} et~al.}{2014}]{si14}
{Singh} K.~P.,  et~al., 2014, in Space Telescopes and Instrumentation 2014:
  Ultraviolet to Gamma Ray. p. 91441S, \mn@doi{10.1117/12.2062667}

\bibitem[\protect\citeauthoryear{{Singh} et~al.,}{{Singh} et~al.}{2016}]{si16}
{Singh} K.~P.,  et~al., 2016, in Space Telescopes and Instrumentation 2016:
  Ultraviolet to Gamma Ray. p. 99051E, \mn@doi{10.1117/12.2235309}

\bibitem[\protect\citeauthoryear{{Singh} et~al.,}{{Singh} et~al.}{2017}]{si17}
{Singh} K.~P.,  et~al., 2017, \mn@doi [Journal of Astrophysics and Astronomy]
  {10.1007/s12036-017-9448-7}, \href
  {https://ui.adsabs.harvard.edu/abs/2017JApA...38...29S} {38, 29}

\bibitem[\protect\citeauthoryear{{Sreehari}, {Nandi}, {Das}, {Agrawal},
  {Mandal}, {Ramadevi}  \& {Katoch}}{{Sreehari} et~al.}{2020}]{sr20}
{Sreehari} H.,  {Nandi} A.,  {Das} S.,  {Agrawal} V.~K.,  {Mandal} S.,
  {Ramadevi} M.~C.,   {Katoch} T.,  2020, \mn@doi [\mnras]
  {10.1093/mnras/staa3135}, \href
  {https://ui.adsabs.harvard.edu/abs/2020MNRAS.499.5891S} {499, 5891}

\bibitem[\protect\citeauthoryear{{Taam}, {Chen}  \& {Swank}}{{Taam}
  et~al.}{1997}]{ta97}
{Taam} R.~E.,  {Chen} X.,   {Swank} J.~H.,  1997, \mn@doi [\apjl]
  {10.1086/310812}, \href
  {https://ui.adsabs.harvard.edu/abs/1997ApJ...485L..83T} {485, L83}

\bibitem[\protect\citeauthoryear{{Vilhu}}{{Vilhu}}{2002}]{Vi02}
{Vilhu} O.,  2002, \mn@doi [\aap] {10.1051/0004-6361:20020610}, \href
  {http://adsabs.harvard.edu/abs/2002A%26A...388..936V} {388, 936}

\bibitem[\protect\citeauthoryear{{Yadav}, {Rao}, {Agrawal}, {Paul}, {Seetha}
  \& {Kasturirangan}}{{Yadav} et~al.}{1999}]{ya99}
{Yadav} J.~S.,  {Rao} A.~R.,  {Agrawal} P.~C.,  {Paul} B.,  {Seetha} S.,
  {Kasturirangan} K.,  1999, \mn@doi [\apj] {10.1086/307225}, \href
  {https://ui.adsabs.harvard.edu/abs/1999ApJ...517..935Y} {517, 935}

\bibitem[\protect\citeauthoryear{{Yadav} et~al.,}{{Yadav}
  et~al.}{2016a}]{ya16b}
{Yadav} J.~S.,  et~al., 2016a, \mn@doi [\apj] {10.3847/0004-637X/833/1/27},
  \href {https://ui.adsabs.harvard.edu/abs/2016ApJ...833...27Y} {833, 27}

\bibitem[\protect\citeauthoryear{{Yadav} et~al.,}{{Yadav}
  et~al.}{2016b}]{ya16a}
{Yadav} J.~S.,  et~al., 2016b, in Space Telescopes and Instrumentation 2016:
  Ultraviolet to Gamma Ray. p. 99051D, \mn@doi{10.1117/12.2231857}

\bibitem[\protect\citeauthoryear{{Yan} et~al.,}{{Yan} et~al.}{2018}]{sh18}
{Yan} S.-P.,  et~al., 2018, \mn@doi [\mnras] {10.1093/mnras/stx2885}, \href
  {https://ui.adsabs.harvard.edu/abs/2018MNRAS.474.1214Y} {474, 1214}

\makeatother
\end{thebibliography}
\label{lastpage}
\end{document}